\documentclass[%
 reprint,
 nobalancelastpage,
 amsmath,amssymb,
prl,
]{revtex4-1}

\usepackage{verbatim}
\usepackage{siunitx}
\usepackage{color}
\usepackage{graphicx}
\usepackage{dcolumn}
\usepackage{bm}

\begin{document}


\title{Field-Tunable Topological Phase Transitions and Spin-Hall Effects in 2D Crystals}

\author{Maxwell Fishman}
 \altaffiliation{fishman.maxwell@gmail.com}
\author{Debdeep Jena$^{1}$}%
\affiliation{%
 $^1$Department of Electrical and Computer Engineering, Cornell University, Ithaca, NY 14850
}%


\date{\today}

\begin{abstract}
As recent additions to the catalog of 2D crystals, silicene and other silicene-class crystals have numerous unique properties currently being investigated and considered for use in novel device applications. In this paper, we investigate electronic and transport properties of silicene in a field effect transistor geometry. We find that the Berry curvature of silicene-class crystals can be continuously tuned by a perpendicular electric field. By direct calculation of the $Z_2$ invariant, we confirm that an electronic phase transition from a topological insulator to a band insulator occurs when the electric field passes a critical value. In a device setting with asymmetric gate voltages, this field-tunable Berry curvature generates a large spin current transverse to the charge current. When the electric field strength surpasses the critical value, the bulk spin current is found to change direction and greatly decrease in magnitude. This finding of a large magnitude, switchable spin current suggests that the silicene family of 2D crystals could be an attractive candidate for field-tunable charge-spin conversion. Such field-tunable phase transitions between topologically distinct phases could be useful for robust qubits as well.
\end{abstract}

\pacs{Valid PACS appear here}
\maketitle

In an effort to overcome the issues that arise as 3D devices approach near atomic scale lengths, a great deal of investigation is being done into the use of 2D materials~\cite{jena2013}. Graphene, the quintessential 2D material, has many unique properties that could make it useful in devices, but preventing current leakage is difficult due to its gapless band structure \cite{Goerbig2011, CastroNeto2009}. Consequently, 2D crystals of the silicene family are receiving significant attention. Silicene has the same honeycomb lattice structure as graphene, but this class of 2D crystals have larger band gaps due to larger atomic radii and buckled lattice structures. Until recently, silicene and related 2D crystals were primarily considered theoretical since monolayers of silicon rapidly decompose in air. However, advances in material growth processes show the potential to stabilize such 2D crystals and use them in devices \cite{Tao2015, Molle2017}.

Silicene can be considered a silicon analog of graphene and has numerous unique properties \cite{Molle2017, Kara2012, Yamada-Takamura2014}. The larger silicon atoms in silicene cause it to have a unique buckled structure and stronger spin orbit coupling (SOC) than graphene. DFT analysis has shown that the SOC opens up a band gap in silicene of $1.55$ meV \cite{Ezawa2012}. The SOC gap is small relative to compounds such as InSb (split-off band gap $\sim0.8$ eV \cite{Winkler2003}), but large enough to allow for measurement of silicene's properties at temperatures of $\sim50\text{ K}$. The properties of silicene can also be generalized to most honeycomb structured 2D materials with broken inversion symmetry \cite{Molle2017, Akinwande2019, Ezawa2015, Chowdhury2016}. Silicene has also been researched for use in ferromagnetic junction spin devices \cite{Kharadi2020, Vargiamidis2014} as well as bilayer silicene in a FET \cite{Ouyang2018}.

In this work, we investigate how monolayer silicene is affected by a perpendicular electric field and how these affects can be exploited in a independent double gate field effect transistor (IDGFET) setting. In an IDGFET, silicene can be placed in band insulator (BI) and quantum spin hall (QSH) states through gate voltage control. A field-tunable spin hall effect (SHE) causes a pure spin current transverse to the drain charge current. The magnitude and direction of this spin current can be switched between two states, depending on whether the silicene channel is in the BI or QSH state. This as well as the magnitude of the spin current are particularly interesting properties of the silicene IDGFET and silicene-class 2D materials may serve as a lightweight alternative material for spintronic devices. 

Due to the buckled structure of silicene, application of an electric field perpendicular to the silicene plane creates a potential difference between the two sublattices. This potential difference exposes many of the interesting properties of silicene and similar 2D crystals. A low energy approximation of a four-band tight binding Hamiltonian is~\cite{Ezawa2012, Liu2011}
\begin{equation}
\mathcal{H}_\eta=\hbar v_f(k_x\tau_x-\eta k_y\tau_y)+\eta\tau_zh_{11}+q\ell F_z\tau_z,
\end{equation}
where $v_f=\frac{\sqrt{3}at}{2\hbar}=5.5\times10^7$ cm/s is the Fermi velocity, $a=3.86~\text{\AA}$ is the lattice constant, $\tau_i$ is the $i$th Pauli sublattice matrix, $\eta=\pm1$ refers to the $K_\pm$ Dirac point, $F_z$ is the vertical electric field strength, $\ell=0.23\text{ \AA}$ is half the vertical distance between the sublattices, and $h_{11}=-\lambda_{SO}\sigma_z-a\lambda_R(k_y\sigma_x-k_x\sigma_y)$. $\sigma_i$ is the Pauli $i$th spin matrix and $\lambda_{SO}=3.9$ meV and $\lambda_R=0.7$ meV are the effective and Rashba SOC constants. 

Diagonalizing the low energy Hamiltonian, the energy dispersion near the Dirac points is
\begin{equation}
E_{\eta,s_z}=\pm\sqrt{\hbar^2v_f^2k^2+\Bigg(q\ell F_z-\eta s_z\sqrt{\lambda_{SO}^2+a^2\lambda_R^2k^2}\Bigg)^2},
\end{equation}
where $s_z=\pm1$ refers to spin up and spin down states and $k^2=k_x^2+k_y^2$.

The energy dispersion near the $K_+$ Dirac point is plotted in Fig. 1(a) and (c). At $F_z=0$ the bands are spin and valley degenerate and there is a band gap of $2\lambda_{SO}\sim7.8$ meV. When $|F_z|$ is increased, spin degeneracy is broken due to the broken inversion symmetry. Near the $K_+$ Dirac point, as $F_z$ increases from zero the band gap between the spin up states decreases while the gap between the spin down states increases. When $F_z$ reaches the critical value $F_c=\lambda_{SO}/q\ell\approx17\text{ mV/\AA}$, spin up energy dispersion becomes gapless. Increasing $F_z$ further re-opens the spin up band gap. When $F_z$ is decreased below zero, the spin down band becomes gapless at $-F_c$. Near the $K_-$ Dirac point, the modified spin states are opposite that of the $K_+$ Dirac point. Therefore, the two notable regimes are $|F_z|<|F_c|$ and $|F_z|>|F_c|$. The transition from one regime to the other occurs when the band gap closes and re-opens, signifying a non-adiabatic change in the material's Hamiltonian and a topological phase change in its eigenvalue spectrum~\cite{Ezawa2012, Kane2005b}.

\begin{figure}[h]
\centering
\includegraphics[
    scale=0.38
    ]{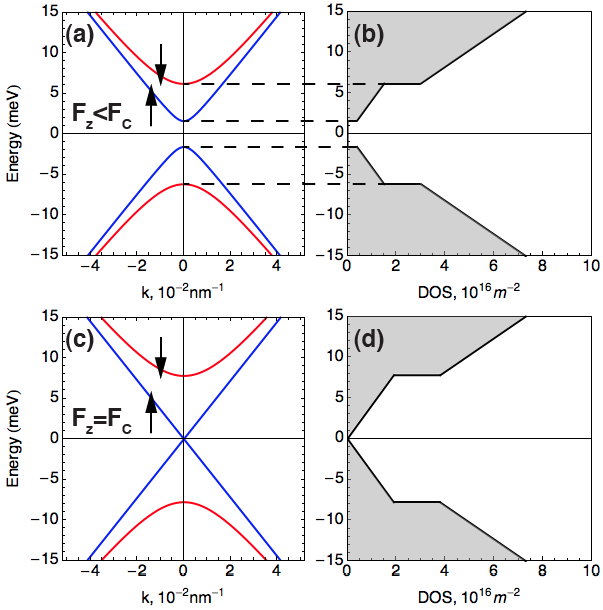}
\caption{\label{fig:BandStruct} Band structure and total density of states (DOS) near the $K_+$ Dirac point. Due to SOC, applying a vertical electric field breaks spin degeneracy into spin up (blue) and spin down (red) bands. (a) Energy dispersion and (b) DOS when the $0<F_z<F_c$. (c) Energy dispersion and (d) DOS when $F_z=F_c$.}
\end{figure}

Relative to effective SOC, $k$-space dependent Rashba SOC has a negligibly small effect on the properties of silicene being investigated in this paper. From this point forward the Rashba spin orbit component will be neglected ($\lambda_R=0$).


The DOS is found by equating the number of states in a infinitesimal volume of $k$-space to that of energy space. Summing over all bands, the total low-energy DOS is
\begin{equation}
\begin{aligned}
g_{total}(E)=\frac{|E|}{\pi\hbar^2v_f^2}\bigg(\Theta\Big(|E|-|eF_z\ell+\lambda_{SO}|\Big)+ \\\Theta\Big(|E|-|eF_z\ell-\lambda_{SO}|\Big)\bigg),
\end{aligned}
\end{equation}

where $\Theta(...)$ is the Heaviside step function.

The total DOS is shown in Fig. 1(b) and (d) for two different electric fields. At $F_z=0$, the bands are spin degenerate, resulting in only one step and a linear DOS. When $F_z\neq0$ two steps occur in the DOS representing the spin split upper and lower bands. At $|F_z|=|F_c|$, the DOS of the lower bands become gapless. As $|F_z|$ increases beyond $|F_c|$, the first step in the DOS re-opens, matching the behavior of the band structure.

Non-trivial Berry curvature is expected in materials with non-zero band gap and broken inversion or time reversal symmetry, which can lead to anomalous hall currents \cite{Xiao2010}. Rather than finding the Berry curvature directly from the Berry potential, a more computationally convenient form is \cite{Xiao2010}
\begin{equation}
\begin{aligned}
\mathbf{\Omega}^n_{\mu\nu}(\textbf{\textit{R}})=i\sum_{n\neq n'}\frac{\langle n | \partial_{{R}^{\mu}} \mathcal{H} | n'\rangle \langle n' | \partial_{{R}^{\nu}} \mathcal{H} |n\rangle}{(\varepsilon_n-\varepsilon_{n'})^2} - \\\frac{\langle n | \partial_{{R}^{\nu}} \mathcal{H} |n'\rangle \langle n' | \partial_{{R}^{\mu}} \mathcal{H} | n\rangle}{(\varepsilon_n-\varepsilon_{n'})^2},
\end{aligned}
\end{equation}
in which $\mathcal{H}$ is the Hamiltonian, $R^{\mu(\nu)}$ is the $\mu\text{th}~(\nu\text{th})$ component of $\textbf{\textit{R}}$ parameter space, which in this case is $k$-space, and $\varepsilon_n$ is the energy dispersion of the $n$th band. The only non-zero Berry curvature component will be $\Omega^n_{xy}$. Therefore, the Berry curvature is locked in the $z$ (out of plane) direction and is found to be
\begin{equation}
\mathbf{\Omega}_{xy}=\frac{\xi\eta\hbar^2v_f^2(qF_z\ell-\eta s_z\lambda_{SO})}{ 2[ \hbar^2k^2v_f^2+(-\eta s_zqF_z\ell+\lambda_{SO})^2 ]^{3/2}}\hat{z},
\end{equation}
where $\xi=\pm1$ signifies the conduction and valence bands.

The conduction band Berry curvature near the $K_+$ Dirac point can be seen in Fig. 2(a) and (b). At $F_z=0$, the Berry curvature for the spin up and spin down states are equal in magnitude, but point in opposite directions. Maximum curvature occurs at the Dirac points. Dependent on the field direction, increasing the field strength causes the curvature of one spin state to increase while the other spin state decreases.

As $F_z$ approaches $F_c$, the spin up Berry curvature in Fig. 2(a) diverges while the spin down curvature continues to decrease. When $F_z$ becomes greater than $F_c$ (Fig. 2(b)), the direction of the spin up Berry curvature changes sign and decreases in magnitude as $F_z$ is increased beyond this point. $F_c$ is the critical field at which the spin up and spin down Berry curvatures transition from pointing in opposite directions to pointing in the same direction. The electron group velocity in a material with a non-trivial Berry curvature is $\textbf{v}_n(\textbf{k})=\frac{\partial \varepsilon_n(\textbf{k})}{\hbar\partial \textbf{k}}-\frac{q}{\hbar}\textbf{F}\times \mathbf{\Omega}_n(\textbf{k})$, where the additional last term is the ``anomalous velocity" \cite{Xiao2010}.
In the presence of a planar electric field, this field-dependent Berry curvature will strongly modify the group velocity and consequently the current density in a silicene-based device. This Berry curvature relationship further indicates a topological phase change, which is evaluated by calculating the $Z_2$ invariant.

\begin{figure}[h]
\centering
\includegraphics[scale=0.19]{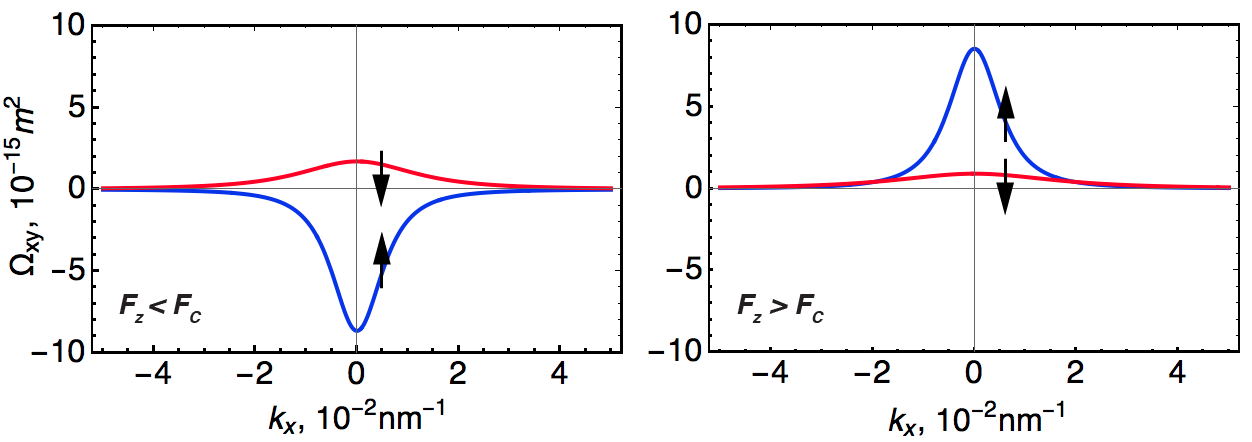}
\caption{\label{fig:BerryCurv} Conduction band Berry curvature near the $K_+$ Dirac point when (a) $0<F_z<F_c$ and (b) $F_z>F_c$. Sweeping $F_z$ through $F_c$ causes the spin up (blue) Berry curvature to change direction, but the spin down (red) Berry curvature does not change direction.}
\end{figure}

It has been reported that silicene is a topological insulator (TI) intrinsically and can be transitioned to a BI by application of a vertical electric field~\cite{Ezawa2012,Ezawa2012b}. Since $s_z$ is conserved in this low energy regime, the $Z_2$ invariant can be evaluated using spin Chern numbers. Integrating the Berry curvatures of each occupied band over the Brillouin zone \cite{Hasan2010, Qi2011, Sheng2006}, the spin Chern numbers are found to be
\begin{equation}
n_{s_z}=\sum_{\eta}\frac{-(\eta qF_z\ell-s_z\lambda_{SO})}{2\sqrt{(-\eta s_zqF_z\ell+\lambda_{SO})^2}},
\end{equation}
where the sum is taken over the Dirac points.

Using the spin Chern number difference, $n_\sigma=(n_\uparrow-n_\downarrow)/2$, the $Z_2$ invariant is found from $\nu=n_\sigma \text{mod}~2$. The Chern invariant is always zero, indicating that silicene is either be a BI or TI. When $|F_z|<|F_c|$, the $Z_2$ invariant is that of a TI ($Z_2=1$). When $|F_z|>|F_c|$, the $Z_2$ invariant changes to that of a BI ($Z_2=0$). As predicted by the closing and re-opening of the band gap, silicene transitions from TI to BI when $F_z$ is swept through $F_c$. This transition point greatly affects bulk transport, which dominates near room temperature.

Using Eq. (2) and (5), the electron group velocity components in the conduction band are found to be
\begin{equation}
v_{i,\eta,s_z}=\frac{\hbar k_iv_f^2}{\sqrt{\hbar^2k^2v_f^2+(-\eta s_zqF_z\ell+\lambda_{SO})^2}}\hat{i},
\end{equation}

\begin{equation}
v_{yBerry,\eta,s_z}=-\frac{\eta qF_x\hbar v_f^2(qF_z\ell-\eta s_z\lambda_{SO})}{2(\hbar^2k^2v_f^2+(-\eta s_zqF_z\ell+\lambda_{SO})^2)^{3/2}}\hat{y},
\end{equation}
where $i=x\text{ or }y$ and $v_{yTotal}=v_{y}+v_{yBerry}$. It is assumed that an electric field in the negative $x$ direction is being applied. The Berry curvature velocity component, $v_{yBerry,\eta,s_z}$, can be recast as 
\begin{equation}
v_{yBerry,\eta,s_z}=-\mu_{B,\eta,s_z}F_x\hat{y},
\end{equation}
where $\mu_B$ is
\begin{equation}
\mu_{B,\eta,s_z}=-\frac{\eta q\hbar v_f^2(qF_z\ell-\eta s_z\lambda_{SO})}{2(\hbar^2k^2v_f^2+(-\eta s_zqF_z\ell+\lambda_{SO})^2)^{3/2}}.
\end{equation}
In this form, $\mu_B$ can be interpreted as a $F_z$-dependent transverse mobility, or ``Berry mobility".


Due to this ``Berry mobility", when $F_x\neq0$ and $F_z\neq0$, there is always an extra non-zero $y$ component in the velocity due to the Berry curvature. This creates asymmetry in the velocity profile and also shifts the zero-velocity point away from the Dirac points. The Berry curvature velocity term rapidly goes to zero as one moves away from the Dirac points. When $|F_z|<|F_c|$ the $y$ direction velocity of spin up and spin down states point in opposite directions and when $|F_z|>|F_c|$ the $y$ direction velocities transition to pointing in the same direction. This predicts a transitioning in the direction and magnitude of a spin Hall effect when silicene changes phase from TI to BI. It should be noted that relativistic effects on electron-electric field dynamics are not included in this analysis.

Using silicene's band structure and velocity profile, a ballistic FET model is built based on the method developed by Natori~\cite{Natori1994}. This method is modified to support the use of silicene as the channel material and the use of independent double gates. A layout of the silicene FET is shown in Fig. 3(a).

Assuming no short channel effects, application of a drain bias will not effect the total carrier density at the source injection point. One can find the source injection point Fermi level using the relationship between the gate voltage and total carrier density and imposing carrier density conservation under a drain bias.

By summing over the energies of the two barriers separately, one finds 
\begin{eqnarray}
q\phi_{b,TG}+qV_{b,TG}-\Delta E_{CB}+(E_f-E_{CB})=qV_{gs,TG} \nonumber \\
q\phi_{b,BG}+qV_{b,BG}-\Delta E_{CB}+(E_f-E_{CB})=qV_{gs,BG}, \nonumber \\
\end{eqnarray}
where $q\phi_b$ is the metal-insulator barrier height, $V_b$ is the voltage drop across the insulator (as seen in Fig 3(b)), $\Delta E_{CB}$ is the insulator-semiconductor conduction band difference, $E_f$ is the semiconductor quasi-Fermi level, and $E_{CB}$ is the conduction band edge. $\Delta E_{CB}$ and $E_{CB}$ are modified by the vertical electric field. Using Gauss' law, $V_b$ is found to be equal to $qn_mt_b/\varepsilon_b\varepsilon_0$. Grouping $\phi_b$ and $\Delta E_{CB}$ into a threshold voltage term, $qV_T=q\phi_b-\Delta E_{CB}$, Eq. (11) are combined into
\begin{equation}
\frac{q^2n_st_{b,TG}}{\varepsilon_{b,TG}\varepsilon_0}+(1+G)(E_f-E_{CB})=q(V_{gs,TG}+GV_{gs,BG}),
\end{equation}
where $G=\frac{\varepsilon_{b,BG}t_{b,TG}}{t_{b,BG}\varepsilon_{b,TG}}$ and the charge neutrality condition, $n_s=n_{m,TG}+n_{m,BG}$, has been used. $n_s$ is the total carrier density in the semiconductor, $t_{b,TG}$ and $t_{b,BG}$ are the top and bottom gate oxide thickness, $\varepsilon_{b,TG}$ and $\varepsilon_{b,BG}$ are the top and bottom gate oxide relative permittivity, and $V_{gs,TG}$ and $V_{gs,BG}$ are the top and bottom gate voltages. It is assumed that the gate voltages are in reference to the threshold voltages.

By summing the occupation function over silicene's conduction bands and applying carrier density conservation, a relationship can be found between $n_s$ and $E_f$ under any drain bias. This relationship is
\begin{equation}
\begin{aligned}
n_{\eta,s_z}=\frac{k_b^2T^2}{\pi v_f^2\hbar^2}\int_{\beta E_{CB}}^{\infty}~du\frac{u}{1+e^{u-\eta_0}}\\=\frac{k_b^2T^2}{2\pi v_f^2\hbar^2}\int_{\beta E_{CB}}^{\infty}~du\Bigg(\frac{u}{1+e^{u-\eta_\rightarrow}}+\frac{u}{1+e^{u-\eta_\leftarrow}}\Bigg),
\end{aligned}
\end{equation}
where $E_{CB}=|qF_z\ell-\eta s_z\lambda_{SO}|$ after modification by the vertical electric field, $\beta=1/k_bT$, and $u=\beta E$. The Fermi level is given by $\eta=\beta E_f$. $\eta_0$ represents the Fermi level without the application of a drain bias and $\eta_\rightarrow$ and $\eta_\leftarrow=\eta_\rightarrow-qV_{ds}$ represent the quasi-Fermi levels of the right and left going carriers when a drain bias is applied. The total carrier density $n_s=\sum_\eta\sum_{s_z}n_{\eta,s_z}=2(n_{+,\uparrow}+n_{+,\downarrow})$ results from the symmetry of the bands. Solving Eq. (13) for $E_f(n_s)$, one can find $n_s(V_{gs,TG},V_{gs,BG})$ using Eq. (12). In order to find $F_z$, the electric field is assumed to linearly change through the channel and the mid-point is taken as the vertical field to which the channel is exposed.

Using $\eta_\rightarrow(V_{ds},V_{gs,TG},V_{gs,BG})$, the total drain charge current can be found using Eq. (7) and (8). The current is found on a per-band basis and the total current is the summation over all bands. Since the only difference between a right-going and left-going carrier is the Fermi level, the total current per conduction band is found to be
\begin{equation}
\begin{aligned}
J_{x,\eta,s_z}=J_0\int_{\beta E_{CB}}^{\infty}du\sqrt{u^2-\beta^2(-\eta s_zqF_z\ell+\lambda_{SO})^2}\\\Bigg(\frac{1}{1+e^{u-\eta_\rightarrow}}-\frac{1}{1+e^{u-\eta_\leftarrow}}\Bigg),
\end{aligned}
\end{equation}
where $J_0=\frac{qk_b^2T^2}{4\pi^2\hbar^2v_f}$, and the total current over all bands is $J_x^{Total}=\sum_\eta\sum_{s_z} J_{\eta,s_z}=2(J_{x,+,\uparrow}+J_{x,+,\downarrow})$.

Fig. 3(c) shows the relationship between current density and symmetric gate voltages at $V_{ds}=0.2\text{ V}$. For asymmetric gate voltages, the current is influenced by the electric field shifting the conduction band edges, but the variations are small. Therefore, the vertical field has little effect on source-drain (longitudinal) transport.

Due to the symmetry of the bands the total longitudinal spin current, $J_{spin}=J_\uparrow-J_\downarrow$, is always zero. Furthermore, it should also be noted that the model developed thus far is for conduction band transport. As experimentally shown~\cite{Tao2015}, once the gate voltages are low enough valence band transport takes over and the current increases. If valence band transport is included, the ON/OFF ratio is reduced at room temperature. Using the symmetry of the conduction and valence bands, the valence band and total current density is shown in Fig. 3(c). In order to increase ON/OFF ratio and reduce leakage, one could operate at lower temperatures (Fig. 3(c) and (d)). Overall, the standard source-drain characteristics of this silicene IDGFET are similar to that of a traditional FET.

\begin{figure}[h]
\centering
\includegraphics[
    height=220px,
    width=236px]{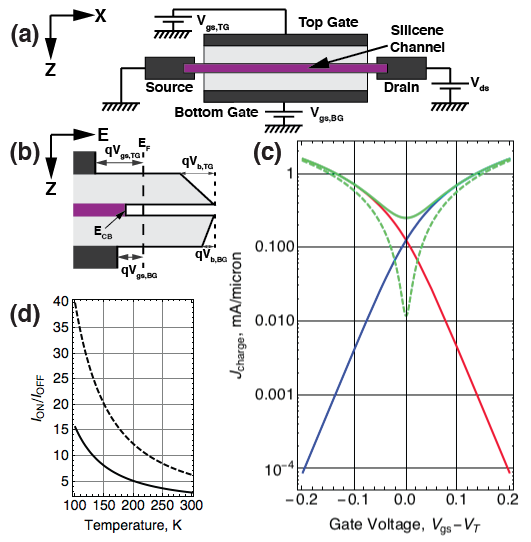}
\caption{\label{fig:FETLayout} Sample layout and drain current density of silicene IDGFET. (a) Sample layout of silicene FET. (b) Band diagram of a IDGFET in the gate-channel-gate direction. (c) Drain current density at 300K, $V_{ds}=0.2\text{ V}$, $t_{b,TG}=t_{b,BG}=2\text{ nm}$, and $\varepsilon_{b,TG}=\varepsilon_{b,BG}=10$ when considering conduction band transport (blue), valence band transport (red), and the total current density (green, solid). Total transport at 50K (green, dashed) is also shown. (d) $I_{ON}/I_{OFF}$ ratio at $V_{gs,TG}=0.1V$ (solid) and $V_{gs,TG}=0.2V$ (dashed). All plots use symmetric gate voltages.}
\end{figure}

Due to the non-trivial Berry curvature, silicene has a current component which is transverse to the longitudinal current. Since there is no voltage applied across the channel in the $y$ direction, the standard group velocity induced current is zero. Therefore, the transverse current is solely a result of silicene's Berry curvature (Fig. 4(a)).

Using the Berry curvature velocity expression, the ballistic current density relation is found to be
\begin{equation}
J_{Berry,\eta,s_z}=J_\Theta\int_{\beta E_{CB}}^{\infty}du\frac{u^{-2}}{1+e^{u-\eta}},
\end{equation}

where $J_\Theta=\frac{-q^2\eta F_x(qF_z\ell-\eta s_z\lambda_{SO})}{4\pi\hbar k_bT}$ and $F_x=V_{ds}/L_{ch}$ is the $x$ direction electric field due to the drain bias. This assumes a linear voltage drop across the channel with $L_{ch}$ being the channel length. This approximation remains valid as long as the transistor is below saturation.

As before, Eq. (15) applies to conduction band transport. Valence band transport can be found using band and Berry curvature symmetries. Using the previously found source injection Fermi level, Fig. 4(b) shows the total transverse spin current at various gate voltages. Due to symmetries of the Berry curvatures, the total Berry charge current, $J_y^{total}=\sum_\eta\sum_{s_z} J_{Berry,\eta,s_z}$, is always zero. The total transverse spin current, $J_y^{spin}=\sum_\eta J_{Berry,\eta,\uparrow}-\sum_\eta J_{Berry,\eta,\downarrow}$, is non-zero and displays a unique switching effect. When $|F_z|<|F_{c}|$ there is a large total spin current. The total spin current switches direction and reduces magnitude when $|F_z|>|F_{c}|$. 

\begin{figure}[h]
\centering
\includegraphics[
    height=170px,
    width=217px]{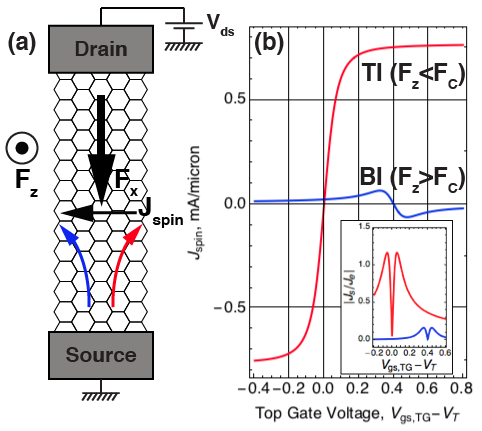}
\caption{\label{fig:SHE} Transverse transport diagram and spin current density. (a) Due to the non-trivial berry curvature, the electric field in the x direction causes carrier transport in the y direction. This transport is spin polarized, creating a transverse spin current. (b) Total spin current density when $|F_z|<|F_c|$ (red) and $|F_z|>|F_c|$ (blue). Inset: spin current ratio, $|J_s/J_e|$, in the TI (red) and BI (blue) regimes. $V_{gs,TG}=V_{gs,BG}$ for the TI regime (red) and $V_{gs,BG}=V_{gs,TG}-0.8V$ for the BI regime (blue). $L_{ch}=10\text{ nm}$ and $V_{ds}=0.1\text{ V}$.}
\end{figure}

Since the spin current is relatively small when $|F_z|>|F_{c}|$, this can be considered as a means to switch the current between a HIGH and LOW state. When the transverse spin current is HIGH the spin current magnitude is comparable to the drain current. An appropriate way to examine this is the spin current ratio, $\theta_{SH}=|J_s/J_e|$, where $J_e$ is the charge drain current density and $J_s$ is the spin current density. Recent values that have been experimentally measured range from $|\theta_{SH}|=0.3$ to $0.4$ in heavy metal based devices \cite{Xu2020, Zhu2020, Zhu2019} and up to 18.62 has been reported in a TI based device \cite{Dc2018}. As seen in the inset of Fig. 4(b), the spin current ratio for silicene reaches a maximum value of $\sim1.15$ at $L_{ch}=10\text{ nm}$. The ratio can be tuned higher if the channel length is reduced or if $V_{ds}$ is increased, but a ratio of $1.15$ at $0.1V$ is quite remarkable. However, one can see from the $\sim1/E_{CB}^2$ form of $v_{yBerry}$ that relativistic effects will be a limiting factor in the maximum spin hall current density as the band gap is reduced. This large spin current magnitude and the switching capability makes silicene an exciting candidate for low power spintronic devices.

In conclusion, when silicene is exposed to an electric field perpendicular to the material plane, spin degeneracy is broken and the Berry curvature and electron group velocity are strongly modified. In an IDGFET setting, silicene's Berry curvature induces a large pure spin current which can be switched between HIGH and LOW states. This spin current is produced without the need for ferromagetic junctions or external magnetic fields. This is particularly surprising considering that heavy elements are usually used to create such spin currents \cite{Liu2012, Xu2020, Zhu2020, Zhu2019}. Sweeping $|F_z|$ through $|F_{c}|$ also causes silicene to transition from TI to BI. This model could lead to a good platform for experimentally investigating relativistic effects on electron dynamics. There have also been theoretical predictions that quantum anomalous hall states occur in response to an exchange field \cite{Ezawa2012b}. Furthermore, there is work on manipulating the electronic and thermal conductivity properties of silicene through mechanical strain \cite{Xie2016, Yan2015}. This, as well as looking at similar properties in other X-ene materials, is a sampling of the numerous areas into which this work could be expanded. This work demonstrates the potential for the use of silicene-class of 2D materials in novel nanoelectronic and spintronics applications. 

\begin{acknowledgments}
We thank Zexuan Zhang for the helpful discussions and proof reading.
\end{acknowledgments}

\bibliography{silicene_SHE_paper}

\end{document}